\documentclass[pra,aps,twocolumn,superscriptaddress,floatfix,showpacs,longbibliography]{revtex4-2}
\usepackage{graphicx}
\usepackage{dcolumn}
\usepackage{bm}
\usepackage{upgreek}
\usepackage{natbib}
\usepackage[normalem]{ulem}
\usepackage{amsmath,amssymb}
\usepackage{dsfont}
\usepackage[english]{babel}
\usepackage{color}

\begin{document}
\title{Metastable supersolid in spin-orbit coupled Bose-Einstein condensates}
\author{Wei-Lei Xia}
\thanks{These authors contributed equally to this work.}
\affiliation{Guangdong Provincial Key Laboratory of Quantum Engineering and
Quantum Materials, School of Physics and Telecommunication Engineering, South
China Normal University, Guangzhou 510006, China}
\affiliation{Guangdong-Hong Kong Joint Laboratory of Quantum Matter, Frontier
Research Institute for Physics, South China Normal University, Guangzhou
510006, China}
\author{Lei Chen}
\thanks{These authors contributed equally to this work.}
\affiliation{Guangdong Provincial Key Laboratory of Quantum Engineering and Quantum Materials, School of Physics and Telecommunication Engineering, South China Normal University, Guangzhou 510006, China}
\affiliation{School of Physics and Electronic
Science, Zunyi Normal University, Zunyi 563006, China}
\author{Tian-Tian Li}
\affiliation{Guangdong Provincial Key Laboratory of Quantum Engineering and
Quantum Materials, School of Physics and Telecommunication Engineering, South
China Normal University, Guangzhou 510006, China}
\affiliation{Guangdong-Hong Kong Joint Laboratory of Quantum Matter, Frontier
Research Institute for Physics, South China Normal University, Guangzhou
510006, China}
\author{Yongping Zhang}
\affiliation{Department of Physics, Shanghai University, Shanghai 200444, China}
\author{Qizhong Zhu}
\email{qzzhu@m.scnu.edu.cn}
\affiliation{Guangdong Provincial Key Laboratory of Quantum Engineering and
Quantum Materials, School of Physics and Telecommunication Engineering, South
China Normal University, Guangzhou 510006, China}
\affiliation{Guangdong-Hong Kong Joint Laboratory of Quantum Matter, Frontier
Research Institute for Physics, South China Normal University, Guangzhou
510006, China}

\date{\today}

\begin{abstract}
Supersolid is a special state of matter with both superfluid properties
and spontaneous modulation of particle density. In this paper, we focus
on the supersolid stripe phase realized in a spin-orbit coupled Bose-Einstein
condensate and explore the properties of a class of metastable supersolids. 
In particular, we study a one-dimensional supersolid whose characteristic
wave number $k$ (magnitude of wave vector) deviates from  $k_{m}$, i.e., the one at ground state. In other words,
the period of density modulation is shorter or longer than the one at ground
state. We find that this class of supersolids can still be stable if 
their wave numbers fall in the range $k_{c1}<k<k_{c2}$, with two thresholds $k_{c1}$ and $k_{c2}$.
Stripes with $k$ outside this range suffer from dynamical instability
with complex Bogoliubov excitation spectrum at long wavelength. 
Experimentally, these stripes with $k$ away from $k_m$ are accessible by exciting the longitudinal spin dipole mode,
resulting in temporal oscillation of stripe period as well as $k$.
Within the mean-field Gross-Pitaevskii
theory, we numerically confirm that for a large enough amplitude of spin dipole oscillation, the stripe states
become unstable through breaking periodicity, in qualitative agreement with the existence of
thresholds of $k$ for stable stripes. Our work extends the concept of supersolid
and uncovers a new class of metastable supersolids to explore.
\end{abstract}

\maketitle

\section{introduction}

Supersolid is an exotic state of matter, which features both superfluid
properties and spontaneous modulation of particle density.
Supersolids
are characterized by two types of gapless Goldstone modes, associated with spontaneous breaking
of U(1) gauge symmetry and spatial translation symmetry, respectively. 
Previously, it has been long
speculated that solid helium-4 at low temperature could be a supersolid \cite{kim_Probable_2004,kim_Observation_2004}, 
which, however,
is overturned by subsequent experiments \cite{day_Lowtemperature_2007,kim_Absence_2012}. In recent years, this concept has stimulated renewed interest
among researchers thanks to significant progress made in the community of cold
atomic gases. Till now, there are three successful schemes to realize a supersolid, i.e., 
the stripe phase of spin-orbit coupled Bose-Einstein condensate (BEC) \cite{wang_SpinOrbit_2010,ho_BoseEinstein_2011,li_Stripe_2017,
bersano_Experimental_2019,putra_Spatial_2020},
BEC coupled with modes of two optical cavities \cite{leonard_Supersolid_2017}, and recently
in dipolar gas with roton excitation spectrum \cite{tanzi_Observation_2019,bottcher_Transient_2019,chomaz_LongLived_2019,
guo_Lowenergy_2019,tanzi_Supersolid_2019,norcia_Twodimensional_2021,petter_Bragg_2021}. 

\begin{figure*}[!htb]
\center{\includegraphics[width=16cm]{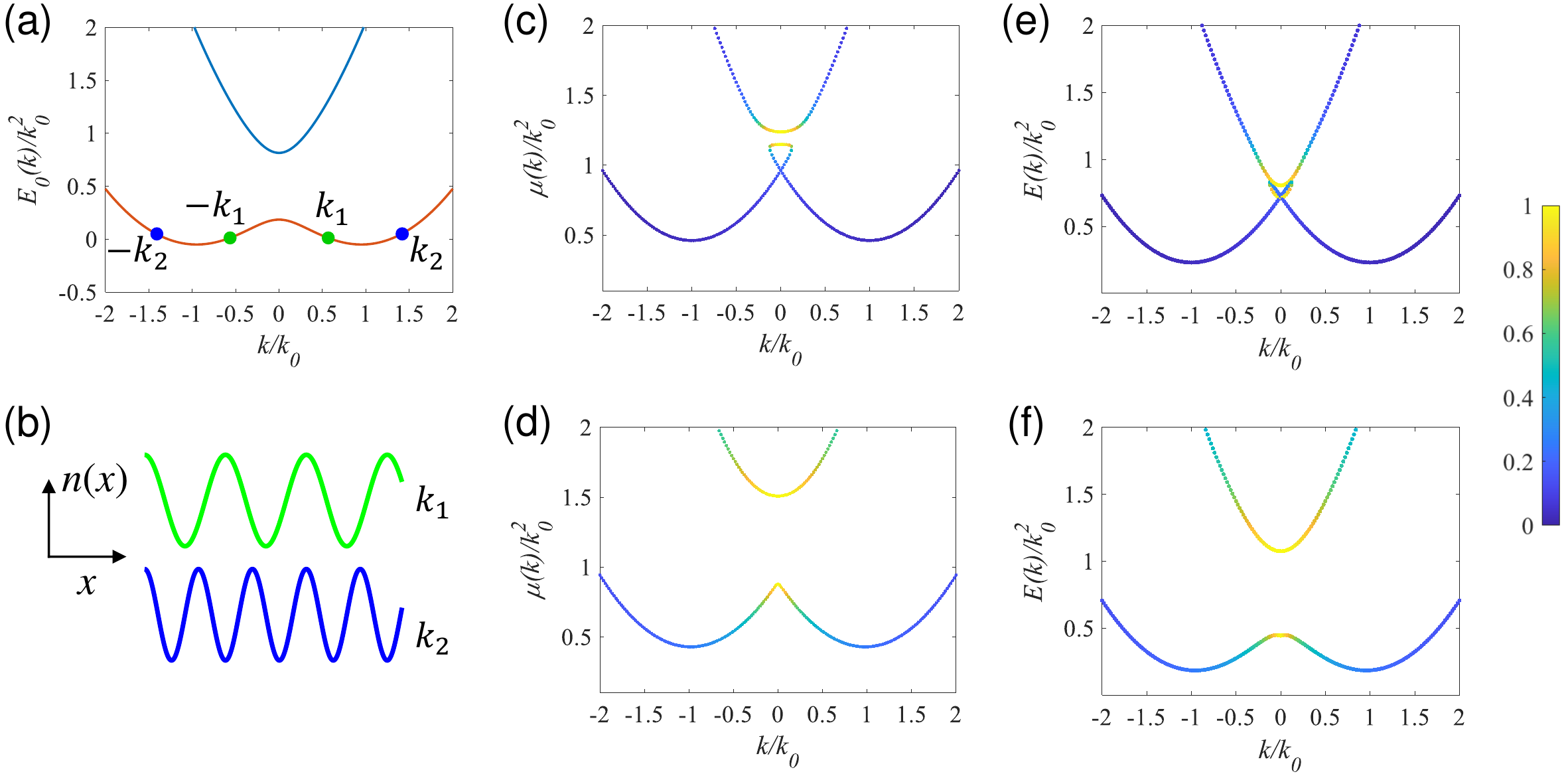}}
\caption{(a) Non-interacting dispersion of spin-orbit coupled BEC. The green (blue) dots denote the momentum
of plane waves with momentum $k_1<k_m$ ($k_2>k_m$), whose interference results in a stripe with period longer (shorter) than
the one at ground state, as illustrated by their density modulations in (b).
(c),(d) Nonlinear dispersions of chemical potential $\mu(k)$ determined by Eq. \ref{muk}. (e),(f) Dispersions of energy $E(k)$ defined in Eq. \ref{ek} corresponding to
(c) and (d), respectively. Line color in (c)-(f) denotes the density contrast $\mathcal{C}$, as shown by the colorbar. 
$\bar{n}g/k_0^2=0.484$, $\bar{n}g_{12}/k_0^2=0.44$, $\Omega/k_0^2=0.09$ for (c),(e), and $\Omega/k_0^2=0.63$ for (a), (d) and (f), respectively. $\bar{n}$
is the average condensate density. \label{fig1}}
\end{figure*}

A supersolid at ground state and its low energy excitations have been extensively
studied, where special attention is paid to the existence of two types of gapless Goldstone modes \cite{li_Superstripes_2013,natale_Excitation_2019,guo_Lowenergy_2019,li_PseudoGoldstone_2021,geier_Exciting_2021}. 
These studies consider the collective excitations in the linear response regime, which reflects the properties
of a supersolid close to ground state.
In fact there are other class of excited states of a supersolid. For example, in the plane wave phase
of spin-orbit coupled BEC, the wave vector does not necessarily coincide with the one at
ground state, but instead can take values away from the lowest energy one.
In particular, these states correspond to an excited state carrying mass current,
and under certain conditions they become metastable superfluid \cite{zhu_Exotic_2012,zheng_Properties_2013,ozawa_Supercurrent_2013}.
Analogously, for the stripe phase, one can also vary the wave vector, 
or equivalently, the period of stripe, to realize a supersolid at excited state (see Figs. \ref{fig1}(a) and \ref{fig1}(b)). 
This scenario is also experimentally relevant as the supersolid prepared may not definitely reach the ground state during limited
observation time.
Will these supersolids be metastable?
If not, what is the underlying instability mechanism? Does there exist a critical spin supercurrent for these metastable supersolids,
in analogy to the critical mass supercurrent in plane wave phase?
These are the interesting questions we aim to answer in this paper.

To be specific, we focus on the supersolid realized in the stripe phase of spin-orbit coupled BEC,
and consider a supersolid whose wave number $k$ (magnitude of wave vector) does not coincide with
the one at ground state, denoted as $k_m$, but instead a supersolid with general $k$.
We find that nonlinear dispersion of chemical potential with respect to $k$ can exhibit a loop structure.
For $k\neq k_m$, the supersolid is
at excited state and may not be stable. We study
the stability of a general stripe state by calculating its Bogoliubov excitation spectrum,
 and find that stripe states with $k$ within the range $k_{c1}<k<k_{c2}$ are metastable. 
 When $k$ falls outside this range, the supersolid
suffers from dynamical instability. These stripes with $k\neq k_m$ can be accessed by exciting the longitudinal spin dipole mode of stripe \cite{geier_Dynamics_2022},
and the thresholds of $k$ are related with the upper limit of the oscillation amplitude of spin dipole mode. 

The rest of this paper is organized as follows. In Sec. \ref{sec2}, we formulate
the spin-orbit coupled BEC within mean-field theory, numerically solve
the wave function of stripe state, and calculate the nonlinear dispersion $\mu(k)$
as a function of wave number $k$. In Sec. \ref{sec3}, we calculate the excitation
spectrum of stripe phase with Bogoliubov theory, from which we determine the stability condition of stripes, and examine the
instability mechanism. We also show that there exists a critical spin supercurrent carried by those stable stripes. 
In Sec. \ref{sec4}, it is demonstrated that stripes with $k\neq k_m$ can be experimentally accessed by exciting the spin dipole mode,
and the stability condition of stripe obtained in Sec. \ref{sec3} is related with
the stability of stripe in response to the 
spin dipole mode excitation within the framework of Gross-Pitaevskii theory.
Finally, we summarize our results and conclude in Sec. \ref{sec5}.

\section{Nonlinear dispersion}
\label{sec2}

Spin-orbit coupled BEC with equal Rashba and Dresselhauss type can be engineered by Raman coupling of
two internal atomic states, as first experimentally realized in NIST group \cite{lin_Spinorbitcoupled_2011}. Assume that the Raman lasers introduce
momentum transfer in $x$ direction, and the trap frequencies in $y$ and $z$
directions are high enough, so that the low energy degree of freedom of this system
can be described by the following one-dimensional effective Hamiltonian with
spin-orbit coupling,
\begin{equation}
\hat{H}_{0}=\frac{\hbar^{2}}{2m}\left(-i\partial_x-k_{0}\sigma_{z}\right)^{2}+\frac{\Omega}{2}\sigma_{x},
\end{equation}
where $k_0$ is the strength of spin-orbit coupling, $\Omega$ is the Raman coupling
strength and Pauli matrices are defined in the pseudospin-$1/2$ Hilbert space. For simplicity, in the following calculations,
we set $\hbar=m=1$. $k_0$ and $k_0^2$ are chosen as momentum and energy units, respectively.
The dispersion of this Hamiltonian has two branches,
as shown in Fig. \ref{fig1}(a).

Now consider Bose-Einstein condensation occurs in this system,
and the two-component condensate can be described by the mean-field
energy functional \cite{li_Quantum_2012},
\begin{align}
{E}\left(\Psi_{\uparrow},\Psi_{\downarrow}\right) = & \int dx\left(\Psi_{\uparrow}^{*},\Psi_{\downarrow}^{*}\right)\hat{H}_{0}\left(\begin{array}{c}
\Psi_{\uparrow}\\
\Psi_{\downarrow}
\end{array}\right)
+g_{\uparrow\downarrow}|\Psi_{\uparrow}|^{2}|\Psi_{\downarrow}|^{2}\nonumber\\
 & +\frac{g}{2}\left(|\Psi_{\uparrow}|^{4}+|\Psi_{\downarrow}|^{4}\right).
 \label{ek}
\end{align}
Here $\Psi_\uparrow$ ($\Psi_\downarrow$) is the condensate wave function of
pseudospin up (down), and $g$ ($g_{12}$) is the interaction strength between same (different) spin species.
The bare values of $g$ and $g_{12}$ are related with the $s$-wave scattering
lengths between atoms of same and different hyperfine states.
For $^{87}$Rb atom, $g$ and $g_{12}$ are very close \cite{lin_Spinorbitcoupled_2011}, making the density contrast of stripe
state very low \cite{li_Quantum_2012}. However, their difference can be enhanced by various schemes, e.g.,
 separating the two components in a bilayer configuration \cite{martone_Approach_2014}.

It is predicted that the ground state
of this model has three phases, with the change of parameters such as $\Omega/k_0^2$, $g$ and $g_{12}$ \cite{li_Quantum_2012}.
For atomic density below a critical value, with the increase of $\Omega/k_0^2$,
the ground state changes from the stripe phase to the plane wave phase,
and then to the zero momentum phase.
The stripe phase spontaneously breaks translation symmetry of the Hamiltonian and
develops periodic modulation of atomic density, which is the focus of this paper.

Minimizing the energy functional with respect to the spinor wave function, one arrives at the stationary Gross-Pitaevskii (GP) equation,
\begin{equation}
\mu\begin{pmatrix}
 \Psi_\uparrow \\
 \Psi_\downarrow  
\end{pmatrix}=\begin{pmatrix}
H_{\uparrow\uparrow} & \frac{\Omega}{2} \\
 \frac{\Omega}{2} & H_{\downarrow\downarrow}
\end{pmatrix}\begin{pmatrix}
\Psi_\uparrow \\
 \Psi_\downarrow  
\end{pmatrix},
\label{muk}
\end{equation}
where $H_{\uparrow\uparrow}={\hbar^2(-i\partial_x-k_0)^2}/{2m}+g|\Psi_\uparrow|^2+g_{12}|\Psi_\downarrow|^2$, and
 $H_{\downarrow\downarrow}={\hbar^2(-i\partial_x+k_0)^2}/{2m}+g|\Psi_\downarrow|^2+g_{12}|\Psi_\uparrow|^2$.
The wave function of a general stripe characterized by $k$ has the following form,
\begin{equation}
\begin{pmatrix}
 \Psi_\uparrow \\
 \Psi_\downarrow  
\end{pmatrix}=\sum_{n=-2L-1}^{2L+1}\begin{pmatrix}
a_n \\
b_n
\end{pmatrix}e^{i n k x},
\end{equation}
where $L$ is a positive integer and the spacing of integer $n$ is $2$.
Note that for the stripe phase, apart from an overall phase factor, the coefficients satisfy the relation
$b_n=a_{-n}^*$, i.e., symmetric under the simultaneous operation of
space inversion and spin flip.
 Since the density now is a spatially periodic function,
the exact stripe wave function should have infinitely many plane waves
similar to Bloch waves in periodic potentials,
although in realistic calculations a finite cutoff $L$ has to be taken.
Plugging the above expansion into the stationary GP equation,
one arrives at a set of nonlinear equations that the coefficients satisfy.
By solving the set of nonlinear equations, one can obtain the nonlinear dispersion,
i.e., $\mu$ as a function of $k$, and also the coefficients of stripe wave function.

Figure \ref{fig1} shows the nonlinear dispersion of both chemical potential $\mu(k)$
and energy ${E}(k)$. It is clear that, at an optimal wave number $k_m$,
the energy of stripe state has the lowest energy.
$k_m$ is close to the one determined in non-interacting
model, i.e., $k_m=k_0\sqrt{1-\Omega^2/4k_0^4}$ \cite{li_Quantum_2012}. 
$k_m$ also corresponds to an optimal stripe period $d_m$, through the
relation $d_m=\pi/k_m$. Away from $k_m$, the energy of stripe state increases,
and a loop or swallowtail structure appears in $\mu(k)$ near $k=0$ for small $\Omega$,
similar to the plane wave case \cite{zhang_Nonlinear_2019}. 
The density contrast of the stripe state, defined as $\mathcal{C}\equiv(n_\mathrm{max}-n_\mathrm{min})/(n_\mathrm{max}+n_\mathrm{min})$ 
 with $n_\mathrm{max}$ ($n_\mathrm{min}$) being local density maximum (minimum), at different $k$ is also indicated by color. One can observe that stripes with small $k$ or long period
has larger density contrast. We also find that the detailed structure of dispersion $\mu(k)$
depends on the cutoff $L$, but the lower branch of dispersion near $k_m$ does not, which is the range of $k$ we are interested in.

Previous studies have only considered the stripe state with this optimal wave number $k_m$.
Here, we have released this constrain and considered a general stripe with
arbitrary $k$, corresponding to ground state or excited
states depending on whether $k=k_{m}$. For $k>k_{m}$, the stripe is effectively
compressed with shorter period; otherwise, it is stretched. Since for $k\neq k_m$, these stripes
are at excited state, it is natural to ask whether these stripes are stable
or not. In other words, does there exist metastable supersolid?
This is the issue we will address in the following section.

\section{Stability of supersolid}
\label{sec3}

The stability of stripes at excited state, or $k\neq k_m$,
can be examined by studying their Bogoliubov excitation spectra.
If the excitation spectrum at all excitation momentum is real and non-negative, these stripes are stable; otherwise, they suffer from
Landau instability or dynamical instability \cite{wu_Superfluidity_2003}.

\begin{figure*}[!htb]
\center{\includegraphics[width=16cm]{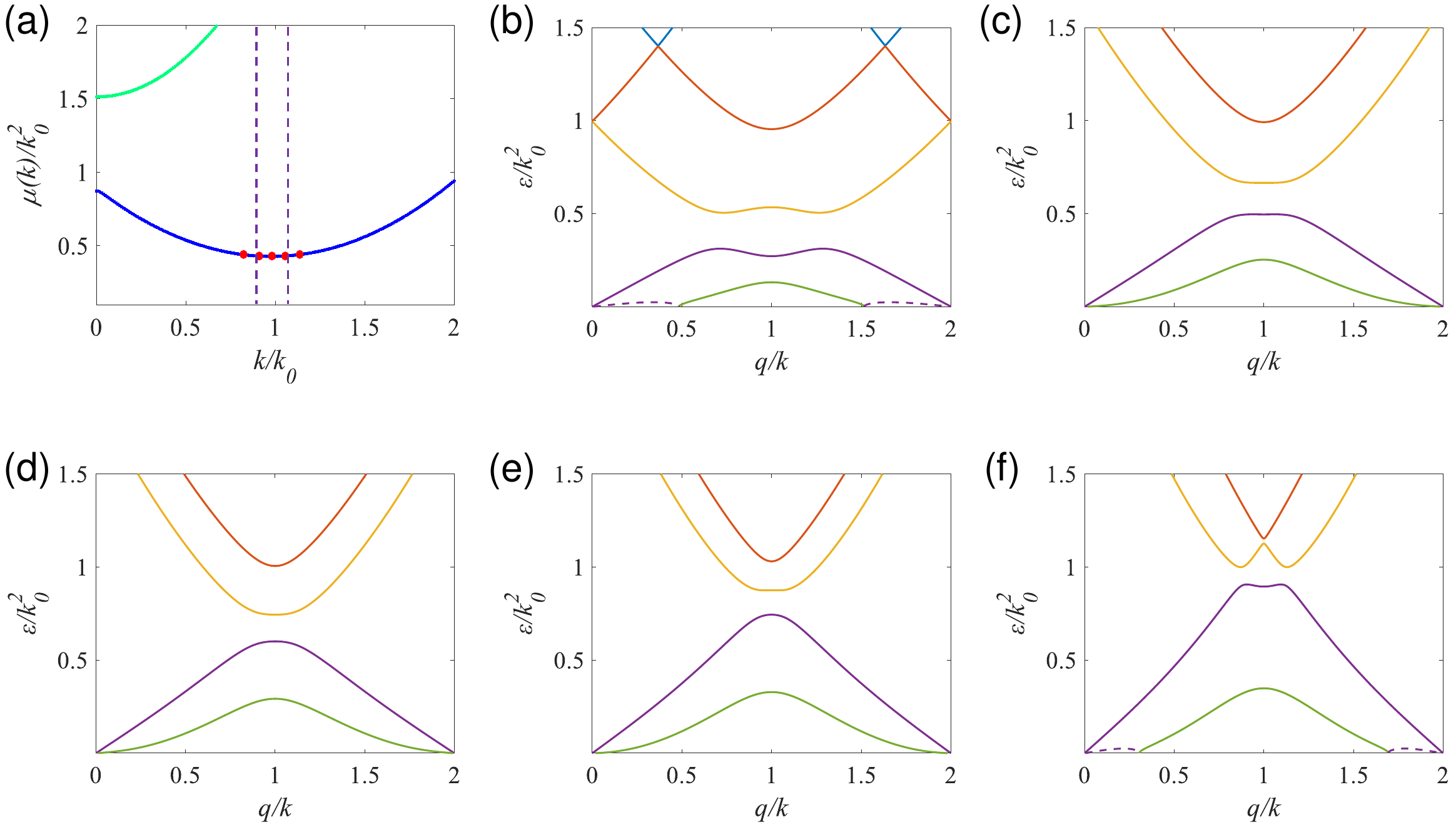}}
\caption{(a) The dispersion $\mu(k)$ with five characteristic momenta $k$, denoted by red dots. From left to right, they correspond to
$k$ in (b)-(f). The dashed lines indicate the lower
and upper thresholds of $k$ of stable stripes. (b)-(f) Bogoliubov excitation spectra
for stripes with five $k$ denoted in (a), corresponding to $k<k_{c1}$, $k_{c1}<k<k_m$, $k=k_m$, $k_m<k<k_{c2}$, and
$k>k_{c2}$, respectively. Excitation spectra in (c)-(e) are all real and non-negative.
Dashed lines in (b) and (f) denote the imaginary parts of complex excitation spectra. $\bar{n}g/k_0^2=0.484$, $\bar{n}g_{12}/k_0^2=0.44$, 
$\Omega/k_0^2=0.59$.  \label{fig2}}
\end{figure*}

We start from the time-dependent GP equation,
\begin{equation}
i\hbar\frac{\partial}{\partial t}\begin{pmatrix}
 \Psi_\uparrow \\
 \Psi_\downarrow  
\end{pmatrix}=\begin{pmatrix}
H_{\uparrow\uparrow} & \frac{\Omega}{2} \\
 \frac{\Omega}{2} & H_{\downarrow\downarrow}
\end{pmatrix}\begin{pmatrix}
\Psi_\uparrow \\
 \Psi_\downarrow  
\end{pmatrix},
\end{equation}
and study the stability of these stripe states with respect to weak perturbations.
Due to periodic modulation of density, the eigenmodes of perturbation 
take the form of Bloch waves,
\begin{align}
\begin{pmatrix}
 \delta\Psi_\uparrow \\
 \delta\Psi_\downarrow  
\end{pmatrix} = & e^{-i\mu(k)t/\hbar}\sum_{n=-2L-1}^{2L+1}\begin{pmatrix}
u_n^\uparrow \\
u_n^\downarrow
\end{pmatrix}e^{i (nk+q)x}e^{-i\varepsilon(q) t/\hbar}\nonumber\\
+ & e^{-i\mu(k)t/\hbar} \sum_{n=-2L-1}^{2L+1}\begin{pmatrix}
v_n^{\uparrow*} \\
v_n^{\downarrow*}
\end{pmatrix}e^{i (nk-q)x}e^{i\varepsilon(q) t/\hbar},
\end{align}
where $u_n^{\uparrow(\downarrow)}$ and $v_n^{\uparrow(\downarrow)}$
are perturbation amplitudes for spin up (down) component.
$q$ is the excitation momentum. Writing $\Psi_{\sigma}'=\Psi_\sigma+\delta\Psi_\sigma$, plugging $\Psi_{\sigma}'$ into the time-dependent GP equation, and only retaining
linear term of perturbation,
one arrives at the Bogoliubov equation that $u_n^{\uparrow(\downarrow)}$ and $v_n^{\uparrow(\downarrow)}$ satisfy, and obtains the excitation spectrum
$\varepsilon(q)$ as well.

The excitation spectra of stripes at different $k$ are depicted in Fig. \ref{fig2}.
Due to periodic density modulation, excitation spectrum $\varepsilon(q)$ has similar feature
with Bloch band of periodic potentials.
For a stripe at ground state, i.e., $k=k_m$, the excitation spectrum features
two branches of real and non-negative gapless excitatons (see Fig. \ref{fig2}(d)), corresponding to
two Goldstone modes associated with spontaneous breaking of U(1) gauge symmetry
and translation symmetry \cite{li_Superstripes_2013}. With the deviation of $k$ from $k_m$, the lower branch of excitation spectrum becomes softer 
(Figs. \ref{fig2}(c) and \ref{fig2}(e)).
When $k<k_{c1}$ or $k>k_{c2}$, with $k_{c1}$ and $k_{c2}$ being the lower and upper threshold of stable stripes,
 the excitation spectrum becomes complex at small $q$,
signalling the dynamical instability of those stripes (Figs. \ref{fig2}(b) and \ref{fig2}(f)). We find that the excitation spectrum always turns complex
starting from $q\rightarrow 0$, indicating that instability first appears in long wavelength excitation. 

\begin{figure}
\centering
\includegraphics[width=0.99\linewidth]{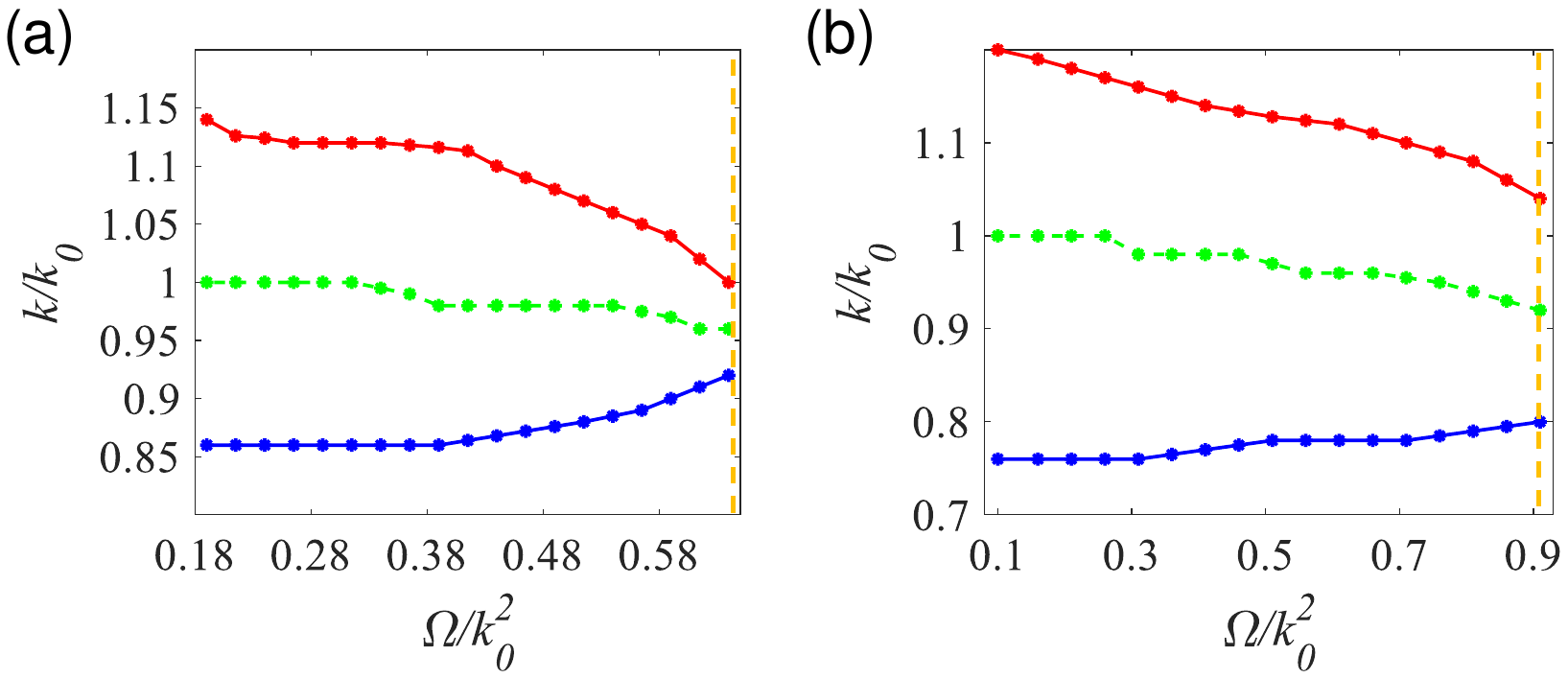}
\caption{The change of upper and lower thresholds $k_{c1}$ and $k_{c2}$ with respect to $\Omega$ are denoted by red and blue lines, respectively.
The change of $k_m$ is also shown in green line for comparison. 
Orange dashed lines indicate the critical value of $\Omega$, beyond which the stripe phase is not ground state. 
$\bar{n}g/k_0^2=0.484$, $\bar{n}g_{12}/k_0^2=0.44$ for (a), and $\bar{n}g/k_0^2=0.57$, $\bar{n}g_{12}/k_0^2=0.456$ for (b). \label{fig3}}
\end{figure}

Both $k_{c1}$ and $k_{c2}$ depend on the Raman coupling strength and the interaction strengths $g$, $g_{12}$.
Figure \ref{fig3} shows the change of two thresholds $k_{c1}$ and $k_{c2}$ as functions of $\Omega/k_0^2$, for two sets of interaction strengths.
One can find that, in both Figs. \ref{fig3}(a) and \ref{fig3}(b), $k_{c1}$ and $k_{c2}$ are getting closer to each other with the increase of $\Omega/k_0^2$,
when approaching the stripe and plane wave phase boundary \cite{li_Quantum_2012}. Besides,
$k_{c1}$ and $k_{c2}$ almost lie symmetrically on opposite sides of $k_m$.
The range of stable $k$ becomes narrower,
consistent with the expectation that the stripe states are more susceptible to dynamical instability when approaching the phase boundary.
Comparison between Figs. \ref{fig3}(a) and \ref{fig3}(b) also shows that, the larger ratio $g/g_{12}$ leads to a wider
range of $k$ of stable stripes.

The stability condition also sets a limit on the magnitude of spin supercurrent \cite{li_Spin_2019}. The stripe state carries a
spin current with the $z$ component of spin current density tensor given by \cite{sun_Definition_2005,zhu_Superfluidity_2015}
\begin{equation}
\boldsymbol{J}_z^s=\frac{1}{2}\left\{\sum_{\sigma=\uparrow,\downarrow}\Psi_\sigma^*(s_z)_{\sigma\sigma}\hat{v}_{\sigma\sigma}
\Psi_\sigma+\mathrm{c.c.}\right\},
\end{equation}
where the velocity operator is defined as $\hat{v}=-i\hbar\partial_x/m-k_0\sigma_z$. For easier illustration, we first take
the stripe wave function with two modes as an example, in which case the spin current reads
\begin{align}
\boldsymbol{J}_z^s= &\frac{k}{2}\left(|a_1|^2-|b_1|^2-|a_{-1}|^2+|b_{-1}|^2\right)-\frac{k_0}{2}\nonumber\\
&-2 k_0 \mathrm{Re}\left[\left(a_1^* a_{-1}+b_1^* b_{-1}\right)\exp(-2 i k x)\right].
\end{align}
After integrating the spin current over a period, the average spin current becomes
\begin{equation}
\boldsymbol{\bar{J}}_z^s=\frac{\pi}{2}\left(|a_1|^2-|b_1|^2-|a_{-1}|^2+|b_{-1}|^2\right)-\frac{\pi k_0}{2 k}.
\end{equation}
For a general stripe wave function with multiple momenta, the average spin current is found to be
\begin{equation}
\boldsymbol{\bar{J}}_z^s=\frac{\pi}{2}\sum_{m=1}^{2L+1}m\left(|a_m|^2-|b_m|^2-|a_{-m}|^2+|b_{-m}|^2\right)-\frac{\pi k_0}{2 k},
\end{equation}
where $m=1,3,\dots,2L+1$.
So the thresholds of $k$ for stable stripes also set a limit on the magnitude of spin supercurrent.
The magnitude of spin supercurrent lies within the range $\boldsymbol{\bar{J}}_{c1}\leq\boldsymbol{\bar{J}}_z^s\leq\boldsymbol{\bar{J}}_{c2}$,
with
$\boldsymbol{\bar{J}}_{c1}={\pi}\sum_{m=1}^{2L+1}m\left(|a_m|^2-|b_m|^2-|a_{-m}|^2+|b_{-m}|^2\right)/2-{\pi k_0}/{2 k_{c1}}$, and
$\boldsymbol{\bar{J}}_{c2}={\pi}\sum_{m=1}^{2L+1}m\left(|a_m|^2-|b_m|^2-|a_{-m}|^2+|b_{-m}|^2\right)/2-{\pi k_0}/{2 k_{c2}}$. 
Note that the coefficients $a_{\pm m}$ and $b_{\pm m}$
are also functions of $k$. 

\section{Experimental observation}
\label{sec4}

A stripe with $k$ away from $k_m$ can be accessed by exciting the spin dipole mode of spin-orbit coupled BEC, which is associated
with the temporal change of stripe period \cite{geier_Dynamics_2022}. 
Analogously, to examine the stability of plane wave phase in spin-orbit coupled BEC,
it has been previously proposed to excite the dipole oscillation of the condensate,
whose center-of-mass momentum sweeps over a finite range around the minimum of dispersion \cite{zhang_Collective_2012,ozawa_Supercurrent_2013}. 

Here the spin dipole mode can be excited by first preparing the stripe at ground state with the perturbation $-m\omega^2 x_0 x\sigma_z$ added to the Hamiltonian, 
and then suddenly
releasing the perturbation and observing the temporal evolution of stripe wave function as well as
 oscillation of density modulation period or $k$ \cite{geier_Dynamics_2022}.
Numerically we first find the ground state of the system with the perturbation $-m\omega^2 x_0 x\sigma_z$ and a harmonic trap
 $V(x)=m\omega^2 x^2/2$ by evolving the GP equation in imaginary time. In this way, the spin up/down component of stripe
 is shifted in opposite directions at ground state. Subsequently, at time $t=0$, the perturbation is released and
 real time evolution of stripe is obtained with the time-dependent GP equation,
\begin{equation}
i\hbar\frac{\partial}{\partial t}\begin{pmatrix}
 \Psi_\uparrow \\
 \Psi_\downarrow  
\end{pmatrix}=\begin{pmatrix}
H_{\uparrow\uparrow} & \frac{\Omega}{2} \\
 \frac{\Omega}{2} & H_{\downarrow\downarrow}
\end{pmatrix}\begin{pmatrix}
\Psi_\uparrow \\
 \Psi_\downarrow  
\end{pmatrix},
\end{equation}
where $H_{\uparrow\uparrow}={\hbar^2(-i\partial_x-k_0)^2}/{2m}+g|\Psi_\uparrow|^2+g_{12}|\Psi_\downarrow|^2+V(x)$, and
$H_{\downarrow\downarrow}={\hbar^2(-i\partial_x+k_0)^2}/{2m}+g|\Psi_\downarrow|^2+g_{12}|\Psi_\uparrow|^2+V(x)$.

\begin{figure}
\centering
\includegraphics[width=0.99\linewidth]{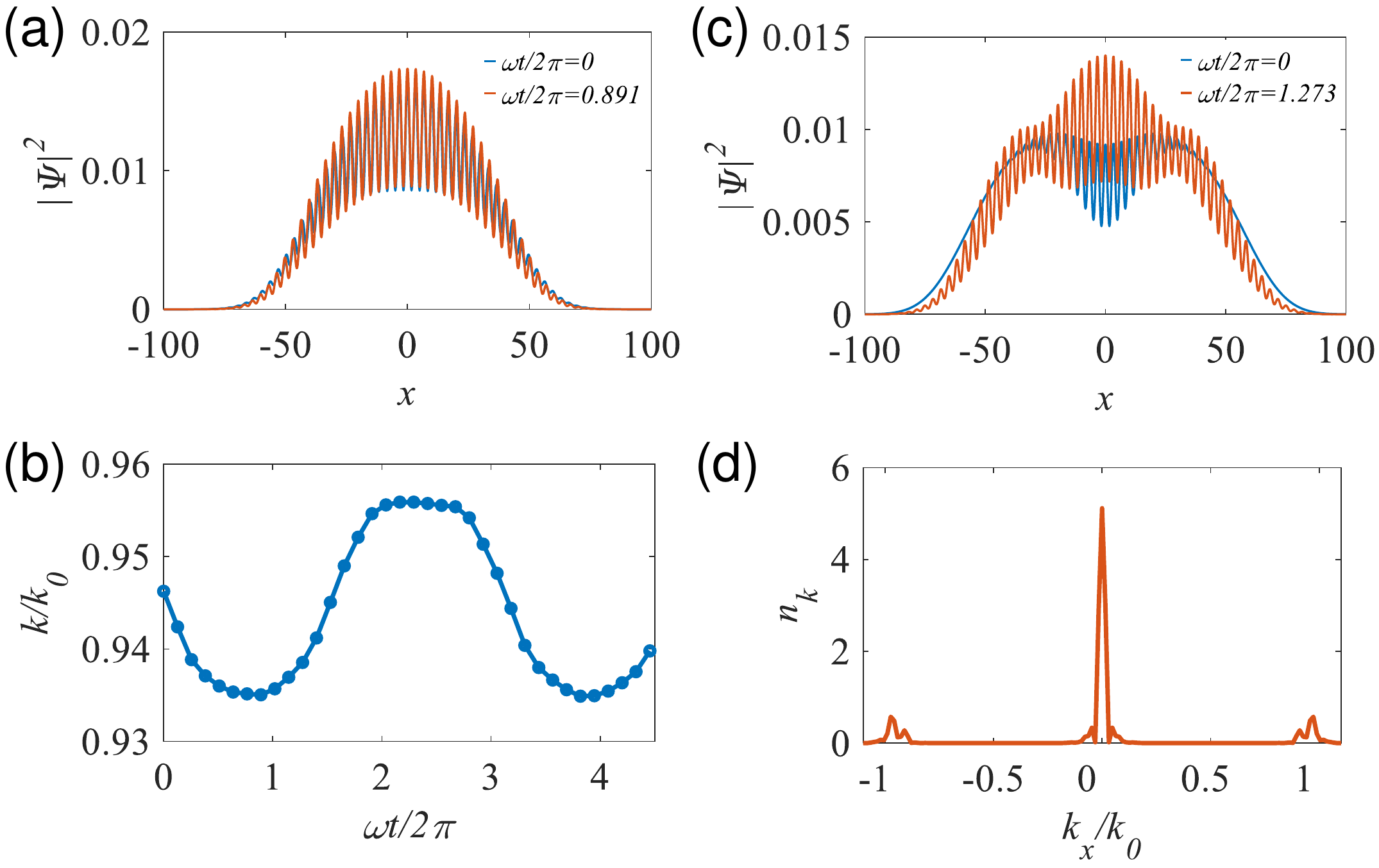}
\caption{(a) Overall density at two moments specified in the inset, for a small initial displacement $x_0=2.5$.
The stripe has well-defined period and $k$ in this case.
The temporal oscillation of $k$ is shown in (b). (c) Overall density for a large initial $x_0=25$, with its momentum distribution at 
$\omega t/2\pi=1.273$ shown in (d). The period and $k$ of stripe are not well-defined in this case. 
Momentum distribution is obtained by Fourier transform of density, with momentum divided by $2$ for comparison
with $k$ defined in this work. The trap frequency is 
$\hbar\omega/k_0^2=0.002$. The other parameters are $\bar{n}g/k_0^2=0.484$, $\bar{n}g_{12}/k_0^2=0.44$ and $\Omega/k_0^2=0.647$,
same as Fig. \ref{fig3}(a). \label{fig4}}
\end{figure}

Figures \ref{fig4}(a) and \ref{fig4}(c) depict the overall density distribution $|\Psi|^2$ of stripe at two different moments for two different initial displacements $x_0$.
Also shown are the momentum distribution obtained by Fourier transform of overall density (Fig. \ref{fig4}(d)), and the oscillation of $k$ 
with time (Fig. \ref{fig4}(b)).
We find that, for a small initial displacement $x_0$, the overall profile of density remains similar and the momentum distribution
always features a peak centred at $k_x=0$ and a pair of peaks centred at $k_x=\pm k$ which is associated with stripe period,
as shown in Figs. \ref{fig4}(a) and \ref{fig4}(b).
In this case, the stripe has well-defined periodicity, and the oscillation amplitude of $k$ around $k_m$ is small, lying within the stable
regime in Fig. \ref{fig3}(a) (see the orange dashed line within stable range). 
The oscillation behaviour of $k$
is consistent with recent studies \cite{geier_Dynamics_2022}. On the other hand, for large initial $x_0$, the momentum distribution
 develops multiple peaks centred at nonzero momentum, which means the crucial feature of stripe is destroyed with no definite period,
 as shown in Figs. \ref{fig4}(c) and \ref{fig4}(d).
This behaviour qualitatively confirms our expectation that, when the oscillation magnitude of $k$ is large for large initial $x_0$,
the stripe can suffer from dynamical instability, with significant change in the feature of density distribution.
So the stability region of $k$ effectively corresponds to a limit on the amplitude of spin dipole oscillation.
Note also that the agreement here is qualitative, as there is a harmonic trap included in the time evolution of
GP equation but not present in the calculation of Bogoliubov excitation spectrum. The agreement should be better
for a weaker trap with harmonic oscillator length $a_x$ much larger than stripe period.

\section{Conclusion and Acknowledgement}
\label{sec5}

In summary, we have studied the properties of a class of metastable supersolid stripes in spin-orbit coupled BEC.
We find that for a stripe with general wave number $k$, the nonlinear dispersion $\mu(k)$ can exhibit a loop structure.
Besides, when the wave number $k$ of stripes deviates from the one at ground state, supersolids can still be
stable with respect to low energy excitations. There exist two thresholds $k_{c1}$ and $k_{c2}$, and for stripes with 
$k_{c1}<k<k_{c2}$, their Bogoliubov excitation spectra are always real and non-negative. For stripes with $k$ outside this
regime, the excitation spectrum at long wavelength becomes complex, signalling the dynamical instability of those stripes.
We also point out that, the thresholds of $k$ for stable stripes are related with the stability criteria of stripes in response
 to spin dipole oscillation, which is accompanied by temporal oscillation of $k$. 
For spin dipole oscillation of large amplitude, the dramatic change
in the feature of stripe wave functions can be qualitatively explained by the instability mechanism we propose here.
With the recent experimental progress on compressional oscillations in dipolar supersolid \cite{tanzi_Supersolid_2019}, it will be
an interesting subject to explore similar stability properties there.

W.X., T.L., and Q.Z. are supported by the National Key Research and Development Program of China (Grant No. 2022YFA1405304), the National Natural
Science Foundation of China (Grant No. 12004118), and the Guangdong Basic and
Applied Basic Research Foundation (Grants No. 2020A1515110228 and No. 2021A1515010212). 
L.C. is supported by the National Natural Science Foundation of China (Grant No. 12264061)
and the Science Foundation of Guizhou Science and Technology Department (Grant No. QKHJZ[2021]033). 
Y.Z. is supported by the National Natural Science Foundation of China with Grants No. 11974235 and
11774219.


\begin{thebibliography}{34}%
\makeatletter
\providecommand \@ifxundefined [1]{%
 \@ifx{#1\undefined}
}%
\providecommand \@ifnum [1]{%
 \ifnum #1\expandafter \@firstoftwo
 \else \expandafter \@secondoftwo
 \fi
}%
\providecommand \@ifx [1]{%
 \ifx #1\expandafter \@firstoftwo
 \else \expandafter \@secondoftwo
 \fi
}%
\providecommand \natexlab [1]{#1}%
\providecommand \enquote  [1]{``#1''}%
\providecommand \bibnamefont  [1]{#1}%
\providecommand \bibfnamefont [1]{#1}%
\providecommand \citenamefont [1]{#1}%
\providecommand \href@noop [0]{\@secondoftwo}%
\providecommand \href [0]{\begingroup \@sanitize@url \@href}%
\providecommand \@href[1]{\@@startlink{#1}\@@href}%
\providecommand \@@href[1]{\endgroup#1\@@endlink}%
\providecommand \@sanitize@url [0]{\catcode `\\12\catcode `\$12\catcode
  `\&12\catcode `\#12\catcode `\^12\catcode `\_12\catcode `\%12\relax}%
\providecommand \@@startlink[1]{}%
\providecommand \@@endlink[0]{}%
\providecommand \url  [0]{\begingroup\@sanitize@url \@url }%
\providecommand \@url [1]{\endgroup\@href {#1}{\urlprefix }}%
\providecommand \urlprefix  [0]{URL }%
\providecommand \Eprint [0]{\href }%
\providecommand \doibase [0]{https://doi.org/}%
\providecommand \selectlanguage [0]{\@gobble}%
\providecommand \bibinfo  [0]{\@secondoftwo}%
\providecommand \bibfield  [0]{\@secondoftwo}%
\providecommand \translation [1]{[#1]}%
\providecommand \BibitemOpen [0]{}%
\providecommand \bibitemStop [0]{}%
\providecommand \bibitemNoStop [0]{.\EOS\space}%
\providecommand \EOS [0]{\spacefactor3000\relax}%
\providecommand \BibitemShut  [1]{\csname bibitem#1\endcsname}%
\let\auto@bib@innerbib\@empty
\bibitem [{\citenamefont {Kim}\ and\ \citenamefont
  {Chan}(2004{\natexlab{a}})}]{kim_Probable_2004}%
  \BibitemOpen
  \bibfield  {author} {\bibinfo {author} {\bibfnamefont {E.}~\bibnamefont
  {Kim}}\ and\ \bibinfo {author} {\bibfnamefont {M.~H.~W.}\ \bibnamefont
  {Chan}},\ }\href@noop {} {\bibfield  {journal} {\bibinfo  {journal} {Nature}\
  }\textbf {\bibinfo {volume} {427}},\ \bibinfo {pages} {225} (\bibinfo {year}
  {2004}{\natexlab{a}})}\BibitemShut {NoStop}%
\bibitem [{\citenamefont {Kim}\ and\ \citenamefont
  {Chan}(2004{\natexlab{b}})}]{kim_Observation_2004}%
  \BibitemOpen
  \bibfield  {author} {\bibinfo {author} {\bibfnamefont {E.}~\bibnamefont
  {Kim}}\ and\ \bibinfo {author} {\bibfnamefont {M.~H.~W.}\ \bibnamefont
  {Chan}},\ }\href@noop {} {\bibfield  {journal} {\bibinfo  {journal}
  {Science}\ }\textbf {\bibinfo {volume} {305}},\ \bibinfo {pages} {1941}
  (\bibinfo {year} {2004}{\natexlab{b}})}\BibitemShut {NoStop}%
\bibitem [{\citenamefont {Day}\ and\ \citenamefont
  {Beamish}(2007)}]{day_Lowtemperature_2007}%
  \BibitemOpen
  \bibfield  {author} {\bibinfo {author} {\bibfnamefont {J.}~\bibnamefont
  {Day}}\ and\ \bibinfo {author} {\bibfnamefont {J.}~\bibnamefont {Beamish}},\
  }\href@noop {} {\bibfield  {journal} {\bibinfo  {journal} {Nature}\ }\textbf
  {\bibinfo {volume} {450}},\ \bibinfo {pages} {853} (\bibinfo {year}
  {2007})}\BibitemShut {NoStop}%
\bibitem [{\citenamefont {Kim}\ and\ \citenamefont
  {Chan}(2012)}]{kim_Absence_2012}%
  \BibitemOpen
  \bibfield  {author} {\bibinfo {author} {\bibfnamefont {D.~Y.}\ \bibnamefont
  {Kim}}\ and\ \bibinfo {author} {\bibfnamefont {M.~H.~W.}\ \bibnamefont
  {Chan}},\ }\href@noop {} {\bibfield  {journal} {\bibinfo  {journal} {Phys.
  Rev. Lett.}\ }\textbf {\bibinfo {volume} {109}},\ \bibinfo {pages} {155301}
  (\bibinfo {year} {2012})}\BibitemShut {NoStop}%
\bibitem [{\citenamefont {Wang}\ \emph {et~al.}(2010)\citenamefont {Wang},
  \citenamefont {Gao}, \citenamefont {Jian},\ and\ \citenamefont
  {Zhai}}]{wang_SpinOrbit_2010}%
  \BibitemOpen
  \bibfield  {author} {\bibinfo {author} {\bibfnamefont {C.}~\bibnamefont
  {Wang}}, \bibinfo {author} {\bibfnamefont {C.}~\bibnamefont {Gao}}, \bibinfo
  {author} {\bibfnamefont {C.-M.}\ \bibnamefont {Jian}},\ and\ \bibinfo
  {author} {\bibfnamefont {H.}~\bibnamefont {Zhai}},\ }\href@noop {} {\bibfield
   {journal} {\bibinfo  {journal} {Phys. Rev. Lett.}\ }\textbf {\bibinfo
  {volume} {105}},\ \bibinfo {pages} {160403} (\bibinfo {year}
  {2010})}\BibitemShut {NoStop}%
\bibitem [{\citenamefont {Ho}\ and\ \citenamefont
  {Zhang}(2011)}]{ho_BoseEinstein_2011}%
  \BibitemOpen
  \bibfield  {author} {\bibinfo {author} {\bibfnamefont {T.-L.}\ \bibnamefont
  {Ho}}\ and\ \bibinfo {author} {\bibfnamefont {S.}~\bibnamefont {Zhang}},\
  }\href@noop {} {\bibfield  {journal} {\bibinfo  {journal} {Phys. Rev. Lett.}\
  }\textbf {\bibinfo {volume} {107}},\ \bibinfo {pages} {150403} (\bibinfo
  {year} {2011})}\BibitemShut {NoStop}%
\bibitem [{\citenamefont {Li}\ \emph {et~al.}(2017)\citenamefont {Li},
  \citenamefont {Lee}, \citenamefont {Huang}, \citenamefont {Burchesky},
  \citenamefont {Shteynas}, \citenamefont {Top}, \citenamefont {Jamison},\ and\
  \citenamefont {Ketterle}}]{li_Stripe_2017}%
  \BibitemOpen
  \bibfield  {author} {\bibinfo {author} {\bibfnamefont {J.-R.}\ \bibnamefont
  {Li}}, \bibinfo {author} {\bibfnamefont {J.}~\bibnamefont {Lee}}, \bibinfo
  {author} {\bibfnamefont {W.}~\bibnamefont {Huang}}, \bibinfo {author}
  {\bibfnamefont {S.}~\bibnamefont {Burchesky}}, \bibinfo {author}
  {\bibfnamefont {B.}~\bibnamefont {Shteynas}}, \bibinfo {author}
  {\bibfnamefont {F.~{\c C}.}\ \bibnamefont {Top}}, \bibinfo {author}
  {\bibfnamefont {A.~O.}\ \bibnamefont {Jamison}},\ and\ \bibinfo {author}
  {\bibfnamefont {W.}~\bibnamefont {Ketterle}},\ }\href@noop {} {\bibfield
  {journal} {\bibinfo  {journal} {Nature}\ }\textbf {\bibinfo {volume} {543}},\
  \bibinfo {pages} {91} (\bibinfo {year} {2017})}\BibitemShut {NoStop}%
\bibitem [{\citenamefont {Bersano}\ \emph {et~al.}(2019)\citenamefont
  {Bersano}, \citenamefont {Hou}, \citenamefont {Mossman}, \citenamefont
  {Gokhroo}, \citenamefont {Luo}, \citenamefont {Sun}, \citenamefont {Zhang},\
  and\ \citenamefont {Engels}}]{bersano_Experimental_2019}%
  \BibitemOpen
  \bibfield  {author} {\bibinfo {author} {\bibfnamefont {T.~M.}\ \bibnamefont
  {Bersano}}, \bibinfo {author} {\bibfnamefont {J.}~\bibnamefont {Hou}},
  \bibinfo {author} {\bibfnamefont {S.}~\bibnamefont {Mossman}}, \bibinfo
  {author} {\bibfnamefont {V.}~\bibnamefont {Gokhroo}}, \bibinfo {author}
  {\bibfnamefont {X.-W.}\ \bibnamefont {Luo}}, \bibinfo {author} {\bibfnamefont
  {K.}~\bibnamefont {Sun}}, \bibinfo {author} {\bibfnamefont {C.}~\bibnamefont
  {Zhang}},\ and\ \bibinfo {author} {\bibfnamefont {P.}~\bibnamefont
  {Engels}},\ }\href@noop {} {\bibfield  {journal} {\bibinfo  {journal} {Phys.
  Rev. A}\ }\textbf {\bibinfo {volume} {99}},\ \bibinfo {pages} {051602(R)}
  (\bibinfo {year} {2019})}\BibitemShut {NoStop}%
\bibitem [{\citenamefont {Putra}\ \emph {et~al.}(2020)\citenamefont {Putra},
  \citenamefont {{Salces-C{\'a}rcoba}}, \citenamefont {Yue}, \citenamefont
  {Sugawa},\ and\ \citenamefont {Spielman}}]{putra_Spatial_2020}%
  \BibitemOpen
  \bibfield  {author} {\bibinfo {author} {\bibfnamefont {A.}~\bibnamefont
  {Putra}}, \bibinfo {author} {\bibfnamefont {F.}~\bibnamefont
  {{Salces-C{\'a}rcoba}}}, \bibinfo {author} {\bibfnamefont {Y.}~\bibnamefont
  {Yue}}, \bibinfo {author} {\bibfnamefont {S.}~\bibnamefont {Sugawa}},\ and\
  \bibinfo {author} {\bibfnamefont {I.~B.}\ \bibnamefont {Spielman}},\
  }\href@noop {} {\bibfield  {journal} {\bibinfo  {journal} {Phys. Rev. Lett.}\
  }\textbf {\bibinfo {volume} {124}},\ \bibinfo {pages} {053605} (\bibinfo
  {year} {2020})}\BibitemShut {NoStop}%
\bibitem [{\citenamefont {L{\'e}onard}\ \emph {et~al.}(2017)\citenamefont
  {L{\'e}onard}, \citenamefont {Morales}, \citenamefont {Zupancic},
  \citenamefont {Esslinger},\ and\ \citenamefont
  {Donner}}]{leonard_Supersolid_2017}%
  \BibitemOpen
  \bibfield  {author} {\bibinfo {author} {\bibfnamefont {J.}~\bibnamefont
  {L{\'e}onard}}, \bibinfo {author} {\bibfnamefont {A.}~\bibnamefont
  {Morales}}, \bibinfo {author} {\bibfnamefont {P.}~\bibnamefont {Zupancic}},
  \bibinfo {author} {\bibfnamefont {T.}~\bibnamefont {Esslinger}},\ and\
  \bibinfo {author} {\bibfnamefont {T.}~\bibnamefont {Donner}},\ }\href@noop {}
  {\bibfield  {journal} {\bibinfo  {journal} {Nature}\ }\textbf {\bibinfo
  {volume} {543}},\ \bibinfo {pages} {87} (\bibinfo {year} {2017})}\BibitemShut
  {NoStop}%
\bibitem [{\citenamefont {Tanzi}\ \emph
  {et~al.}(2019{\natexlab{a}})\citenamefont {Tanzi}, \citenamefont {Lucioni},
  \citenamefont {Fam{\`a}}, \citenamefont {Catani}, \citenamefont {Fioretti},
  \citenamefont {Gabbanini}, \citenamefont {Bisset}, \citenamefont {Santos},\
  and\ \citenamefont {Modugno}}]{tanzi_Observation_2019}%
  \BibitemOpen
  \bibfield  {author} {\bibinfo {author} {\bibfnamefont {L.}~\bibnamefont
  {Tanzi}}, \bibinfo {author} {\bibfnamefont {E.}~\bibnamefont {Lucioni}},
  \bibinfo {author} {\bibfnamefont {F.}~\bibnamefont {Fam{\`a}}}, \bibinfo
  {author} {\bibfnamefont {J.}~\bibnamefont {Catani}}, \bibinfo {author}
  {\bibfnamefont {A.}~\bibnamefont {Fioretti}}, \bibinfo {author}
  {\bibfnamefont {C.}~\bibnamefont {Gabbanini}}, \bibinfo {author}
  {\bibfnamefont {R.~N.}\ \bibnamefont {Bisset}}, \bibinfo {author}
  {\bibfnamefont {L.}~\bibnamefont {Santos}},\ and\ \bibinfo {author}
  {\bibfnamefont {G.}~\bibnamefont {Modugno}},\ }\href@noop {} {\bibfield
  {journal} {\bibinfo  {journal} {Phys. Rev. Lett.}\ }\textbf {\bibinfo
  {volume} {122}},\ \bibinfo {pages} {130405} (\bibinfo {year}
  {2019}{\natexlab{a}})}\BibitemShut {NoStop}%
\bibitem [{\citenamefont {B{\"o}ttcher}\ \emph {et~al.}(2019)\citenamefont
  {B{\"o}ttcher}, \citenamefont {Schmidt}, \citenamefont {Wenzel},
  \citenamefont {Hertkorn}, \citenamefont {Guo}, \citenamefont {Langen},\ and\
  \citenamefont {Pfau}}]{bottcher_Transient_2019}%
  \BibitemOpen
  \bibfield  {author} {\bibinfo {author} {\bibfnamefont {F.}~\bibnamefont
  {B{\"o}ttcher}}, \bibinfo {author} {\bibfnamefont {J.-N.}\ \bibnamefont
  {Schmidt}}, \bibinfo {author} {\bibfnamefont {M.}~\bibnamefont {Wenzel}},
  \bibinfo {author} {\bibfnamefont {J.}~\bibnamefont {Hertkorn}}, \bibinfo
  {author} {\bibfnamefont {M.}~\bibnamefont {Guo}}, \bibinfo {author}
  {\bibfnamefont {T.}~\bibnamefont {Langen}},\ and\ \bibinfo {author}
  {\bibfnamefont {T.}~\bibnamefont {Pfau}},\ }\href@noop {} {\bibfield
  {journal} {\bibinfo  {journal} {Phys. Rev. X}\ }\textbf {\bibinfo {volume}
  {9}},\ \bibinfo {pages} {011051} (\bibinfo {year} {2019})}\BibitemShut
  {NoStop}%
\bibitem [{\citenamefont {Chomaz}\ \emph {et~al.}(2019)\citenamefont {Chomaz},
  \citenamefont {Petter}, \citenamefont {Ilzh{\"o}fer}, \citenamefont {Natale},
  \citenamefont {Trautmann}, \citenamefont {Politi}, \citenamefont
  {Durastante}, \citenamefont {{van Bijnen}}, \citenamefont {Patscheider},
  \citenamefont {Sohmen}, \citenamefont {Mark},\ and\ \citenamefont
  {Ferlaino}}]{chomaz_LongLived_2019}%
  \BibitemOpen
  \bibfield  {author} {\bibinfo {author} {\bibfnamefont {L.}~\bibnamefont
  {Chomaz}}, \bibinfo {author} {\bibfnamefont {D.}~\bibnamefont {Petter}},
  \bibinfo {author} {\bibfnamefont {P.}~\bibnamefont {Ilzh{\"o}fer}}, \bibinfo
  {author} {\bibfnamefont {G.}~\bibnamefont {Natale}}, \bibinfo {author}
  {\bibfnamefont {A.}~\bibnamefont {Trautmann}}, \bibinfo {author}
  {\bibfnamefont {C.}~\bibnamefont {Politi}}, \bibinfo {author} {\bibfnamefont
  {G.}~\bibnamefont {Durastante}}, \bibinfo {author} {\bibfnamefont {R.~M.~W.}\
  \bibnamefont {{van Bijnen}}}, \bibinfo {author} {\bibfnamefont
  {A.}~\bibnamefont {Patscheider}}, \bibinfo {author} {\bibfnamefont
  {M.}~\bibnamefont {Sohmen}}, \bibinfo {author} {\bibfnamefont {M.~J.}\
  \bibnamefont {Mark}},\ and\ \bibinfo {author} {\bibfnamefont
  {F.}~\bibnamefont {Ferlaino}},\ }\href@noop {} {\bibfield  {journal}
  {\bibinfo  {journal} {Phys. Rev. X}\ }\textbf {\bibinfo {volume} {9}},\
  \bibinfo {pages} {021012} (\bibinfo {year} {2019})}\BibitemShut {NoStop}%
\bibitem [{\citenamefont {Guo}\ \emph {et~al.}(2019)\citenamefont {Guo},
  \citenamefont {B{\"o}ttcher}, \citenamefont {Hertkorn}, \citenamefont
  {Schmidt}, \citenamefont {Wenzel}, \citenamefont {B{\"u}chler}, \citenamefont
  {Langen},\ and\ \citenamefont {Pfau}}]{guo_Lowenergy_2019}%
  \BibitemOpen
  \bibfield  {author} {\bibinfo {author} {\bibfnamefont {M.}~\bibnamefont
  {Guo}}, \bibinfo {author} {\bibfnamefont {F.}~\bibnamefont {B{\"o}ttcher}},
  \bibinfo {author} {\bibfnamefont {J.}~\bibnamefont {Hertkorn}}, \bibinfo
  {author} {\bibfnamefont {J.-N.}\ \bibnamefont {Schmidt}}, \bibinfo {author}
  {\bibfnamefont {M.}~\bibnamefont {Wenzel}}, \bibinfo {author} {\bibfnamefont
  {H.~P.}\ \bibnamefont {B{\"u}chler}}, \bibinfo {author} {\bibfnamefont
  {T.}~\bibnamefont {Langen}},\ and\ \bibinfo {author} {\bibfnamefont
  {T.}~\bibnamefont {Pfau}},\ }\href@noop {} {\bibfield  {journal} {\bibinfo
  {journal} {Nature}\ }\textbf {\bibinfo {volume} {574}},\ \bibinfo {pages}
  {386} (\bibinfo {year} {2019})}\BibitemShut {NoStop}%
\bibitem [{\citenamefont {Tanzi}\ \emph
  {et~al.}(2019{\natexlab{b}})\citenamefont {Tanzi}, \citenamefont {Roccuzzo},
  \citenamefont {Lucioni}, \citenamefont {Fam{\`a}}, \citenamefont {Fioretti},
  \citenamefont {Gabbanini}, \citenamefont {Modugno}, \citenamefont {Recati},\
  and\ \citenamefont {Stringari}}]{tanzi_Supersolid_2019}%
  \BibitemOpen
  \bibfield  {author} {\bibinfo {author} {\bibfnamefont {L.}~\bibnamefont
  {Tanzi}}, \bibinfo {author} {\bibfnamefont {S.~M.}\ \bibnamefont {Roccuzzo}},
  \bibinfo {author} {\bibfnamefont {E.}~\bibnamefont {Lucioni}}, \bibinfo
  {author} {\bibfnamefont {F.}~\bibnamefont {Fam{\`a}}}, \bibinfo {author}
  {\bibfnamefont {A.}~\bibnamefont {Fioretti}}, \bibinfo {author}
  {\bibfnamefont {C.}~\bibnamefont {Gabbanini}}, \bibinfo {author}
  {\bibfnamefont {G.}~\bibnamefont {Modugno}}, \bibinfo {author} {\bibfnamefont
  {A.}~\bibnamefont {Recati}},\ and\ \bibinfo {author} {\bibfnamefont
  {S.}~\bibnamefont {Stringari}},\ }\href@noop {} {\bibfield  {journal}
  {\bibinfo  {journal} {Nature}\ }\textbf {\bibinfo {volume} {574}},\ \bibinfo
  {pages} {382} (\bibinfo {year} {2019}{\natexlab{b}})}\BibitemShut {NoStop}%
\bibitem [{\citenamefont {Norcia}\ \emph {et~al.}(2021)\citenamefont {Norcia},
  \citenamefont {Politi}, \citenamefont {Klaus}, \citenamefont {Poli},
  \citenamefont {Sohmen}, \citenamefont {Mark}, \citenamefont {Bisset},
  \citenamefont {Santos},\ and\ \citenamefont
  {Ferlaino}}]{norcia_Twodimensional_2021}%
  \BibitemOpen
  \bibfield  {author} {\bibinfo {author} {\bibfnamefont {M.~A.}\ \bibnamefont
  {Norcia}}, \bibinfo {author} {\bibfnamefont {C.}~\bibnamefont {Politi}},
  \bibinfo {author} {\bibfnamefont {L.}~\bibnamefont {Klaus}}, \bibinfo
  {author} {\bibfnamefont {E.}~\bibnamefont {Poli}}, \bibinfo {author}
  {\bibfnamefont {M.}~\bibnamefont {Sohmen}}, \bibinfo {author} {\bibfnamefont
  {M.~J.}\ \bibnamefont {Mark}}, \bibinfo {author} {\bibfnamefont {R.~N.}\
  \bibnamefont {Bisset}}, \bibinfo {author} {\bibfnamefont {L.}~\bibnamefont
  {Santos}},\ and\ \bibinfo {author} {\bibfnamefont {F.}~\bibnamefont
  {Ferlaino}},\ }\href@noop {} {\bibfield  {journal} {\bibinfo  {journal}
  {Nature}\ }\textbf {\bibinfo {volume} {596}},\ \bibinfo {pages} {357}
  (\bibinfo {year} {2021})}\BibitemShut {NoStop}%
\bibitem [{\citenamefont {Petter}\ \emph {et~al.}(2021)\citenamefont {Petter},
  \citenamefont {Patscheider}, \citenamefont {Natale}, \citenamefont {Mark},
  \citenamefont {Baranov}, \citenamefont {{van Bijnen}}, \citenamefont
  {Roccuzzo}, \citenamefont {Recati}, \citenamefont {Blakie}, \citenamefont
  {Baillie}, \citenamefont {Chomaz},\ and\ \citenamefont
  {Ferlaino}}]{petter_Bragg_2021}%
  \BibitemOpen
  \bibfield  {author} {\bibinfo {author} {\bibfnamefont {D.}~\bibnamefont
  {Petter}}, \bibinfo {author} {\bibfnamefont {A.}~\bibnamefont {Patscheider}},
  \bibinfo {author} {\bibfnamefont {G.}~\bibnamefont {Natale}}, \bibinfo
  {author} {\bibfnamefont {M.~J.}\ \bibnamefont {Mark}}, \bibinfo {author}
  {\bibfnamefont {M.~A.}\ \bibnamefont {Baranov}}, \bibinfo {author}
  {\bibfnamefont {R.}~\bibnamefont {{van Bijnen}}}, \bibinfo {author}
  {\bibfnamefont {S.~M.}\ \bibnamefont {Roccuzzo}}, \bibinfo {author}
  {\bibfnamefont {A.}~\bibnamefont {Recati}}, \bibinfo {author} {\bibfnamefont
  {B.}~\bibnamefont {Blakie}}, \bibinfo {author} {\bibfnamefont
  {D.}~\bibnamefont {Baillie}}, \bibinfo {author} {\bibfnamefont
  {L.}~\bibnamefont {Chomaz}},\ and\ \bibinfo {author} {\bibfnamefont
  {F.}~\bibnamefont {Ferlaino}},\ }\href@noop {} {\bibfield  {journal}
  {\bibinfo  {journal} {Phys. Rev. A}\ }\textbf {\bibinfo {volume} {104}},\
  \bibinfo {pages} {L011302} (\bibinfo {year} {2021})}\BibitemShut {NoStop}%
\bibitem [{\citenamefont {Li}\ \emph {et~al.}(2013)\citenamefont {Li},
  \citenamefont {Martone}, \citenamefont {Pitaevskii},\ and\ \citenamefont
  {Stringari}}]{li_Superstripes_2013}%
  \BibitemOpen
  \bibfield  {author} {\bibinfo {author} {\bibfnamefont {Y.}~\bibnamefont
  {Li}}, \bibinfo {author} {\bibfnamefont {G.~I.}\ \bibnamefont {Martone}},
  \bibinfo {author} {\bibfnamefont {L.~P.}\ \bibnamefont {Pitaevskii}},\ and\
  \bibinfo {author} {\bibfnamefont {S.}~\bibnamefont {Stringari}},\ }\href@noop
  {} {\bibfield  {journal} {\bibinfo  {journal} {Phys. Rev. Lett.}\ }\textbf
  {\bibinfo {volume} {110}},\ \bibinfo {pages} {235302} (\bibinfo {year}
  {2013})}\BibitemShut {NoStop}%
\bibitem [{\citenamefont {Natale}\ \emph {et~al.}(2019)\citenamefont {Natale},
  \citenamefont {{van Bijnen}}, \citenamefont {Patscheider}, \citenamefont
  {Petter}, \citenamefont {Mark}, \citenamefont {Chomaz},\ and\ \citenamefont
  {Ferlaino}}]{natale_Excitation_2019}%
  \BibitemOpen
  \bibfield  {author} {\bibinfo {author} {\bibfnamefont {G.}~\bibnamefont
  {Natale}}, \bibinfo {author} {\bibfnamefont {R.~M.~W.}\ \bibnamefont {{van
  Bijnen}}}, \bibinfo {author} {\bibfnamefont {A.}~\bibnamefont {Patscheider}},
  \bibinfo {author} {\bibfnamefont {D.}~\bibnamefont {Petter}}, \bibinfo
  {author} {\bibfnamefont {M.~J.}\ \bibnamefont {Mark}}, \bibinfo {author}
  {\bibfnamefont {L.}~\bibnamefont {Chomaz}},\ and\ \bibinfo {author}
  {\bibfnamefont {F.}~\bibnamefont {Ferlaino}},\ }\href@noop {} {\bibfield
  {journal} {\bibinfo  {journal} {Phys. Rev. Lett.}\ }\textbf {\bibinfo
  {volume} {123}},\ \bibinfo {pages} {050402} (\bibinfo {year}
  {2019})}\BibitemShut {NoStop}%
\bibitem [{\citenamefont {Li}\ \emph {et~al.}(2021)\citenamefont {Li},
  \citenamefont {Luo}, \citenamefont {Hou},\ and\ \citenamefont
  {Zhang}}]{li_PseudoGoldstone_2021}%
  \BibitemOpen
  \bibfield  {author} {\bibinfo {author} {\bibfnamefont {G.-Q.}\ \bibnamefont
  {Li}}, \bibinfo {author} {\bibfnamefont {X.-W.}\ \bibnamefont {Luo}},
  \bibinfo {author} {\bibfnamefont {J.}~\bibnamefont {Hou}},\ and\ \bibinfo
  {author} {\bibfnamefont {C.}~\bibnamefont {Zhang}},\ }\href@noop {}
  {\bibfield  {journal} {\bibinfo  {journal} {Phys. Rev. A}\ }\textbf {\bibinfo
  {volume} {104}},\ \bibinfo {pages} {023311} (\bibinfo {year}
  {2021})}\BibitemShut {NoStop}%
\bibitem [{\citenamefont {Geier}\ \emph {et~al.}(2021)\citenamefont {Geier},
  \citenamefont {Martone}, \citenamefont {Hauke},\ and\ \citenamefont
  {Stringari}}]{geier_Exciting_2021}%
  \BibitemOpen
  \bibfield  {author} {\bibinfo {author} {\bibfnamefont {K.~T.}\ \bibnamefont
  {Geier}}, \bibinfo {author} {\bibfnamefont {G.~I.}\ \bibnamefont {Martone}},
  \bibinfo {author} {\bibfnamefont {P.}~\bibnamefont {Hauke}},\ and\ \bibinfo
  {author} {\bibfnamefont {S.}~\bibnamefont {Stringari}},\ }\href@noop {}
  {\bibfield  {journal} {\bibinfo  {journal} {Phys. Rev. Lett.}\ }\textbf
  {\bibinfo {volume} {127}},\ \bibinfo {pages} {115301} (\bibinfo {year}
  {2021})}\BibitemShut {NoStop}%
\bibitem [{\citenamefont {Zhu}\ \emph {et~al.}(2012)\citenamefont {Zhu},
  \citenamefont {Zhang},\ and\ \citenamefont {Wu}}]{zhu_Exotic_2012}%
  \BibitemOpen
  \bibfield  {author} {\bibinfo {author} {\bibfnamefont {Q.}~\bibnamefont
  {Zhu}}, \bibinfo {author} {\bibfnamefont {C.}~\bibnamefont {Zhang}},\ and\
  \bibinfo {author} {\bibfnamefont {B.}~\bibnamefont {Wu}},\ }\href@noop {}
  {\bibfield  {journal} {\bibinfo  {journal} {EPL}\ }\textbf {\bibinfo {volume}
  {100}},\ \bibinfo {pages} {50003} (\bibinfo {year} {2012})}\BibitemShut
  {NoStop}%
\bibitem [{\citenamefont {Zheng}\ \emph {et~al.}(2013)\citenamefont {Zheng},
  \citenamefont {Yu}, \citenamefont {Cui},\ and\ \citenamefont
  {Zhai}}]{zheng_Properties_2013}%
  \BibitemOpen
  \bibfield  {author} {\bibinfo {author} {\bibfnamefont {W.}~\bibnamefont
  {Zheng}}, \bibinfo {author} {\bibfnamefont {Z.-Q.}\ \bibnamefont {Yu}},
  \bibinfo {author} {\bibfnamefont {X.}~\bibnamefont {Cui}},\ and\ \bibinfo
  {author} {\bibfnamefont {H.}~\bibnamefont {Zhai}},\ }\href@noop {} {\bibfield
   {journal} {\bibinfo  {journal} {J. Phys. B}\ }\textbf {\bibinfo {volume}
  {46}},\ \bibinfo {pages} {134007} (\bibinfo {year} {2013})}\BibitemShut
  {NoStop}%
\bibitem [{\citenamefont {Ozawa}\ \emph {et~al.}(2013)\citenamefont {Ozawa},
  \citenamefont {Pitaevskii},\ and\ \citenamefont
  {Stringari}}]{ozawa_Supercurrent_2013}%
  \BibitemOpen
  \bibfield  {author} {\bibinfo {author} {\bibfnamefont {T.}~\bibnamefont
  {Ozawa}}, \bibinfo {author} {\bibfnamefont {L.~P.}\ \bibnamefont
  {Pitaevskii}},\ and\ \bibinfo {author} {\bibfnamefont {S.}~\bibnamefont
  {Stringari}},\ }\href@noop {} {\bibfield  {journal} {\bibinfo  {journal}
  {Phys. Rev. A}\ }\textbf {\bibinfo {volume} {87}},\ \bibinfo {pages} {063610}
  (\bibinfo {year} {2013})}\BibitemShut {NoStop}%
\bibitem [{\citenamefont {Geier}\ \emph {et~al.}(2022)\citenamefont {Geier},
  \citenamefont {Martone}, \citenamefont {Hauke}, \citenamefont {Ketterle},\
  and\ \citenamefont {Stringari}}]{geier_Dynamics_2022}%
  \BibitemOpen
  \bibfield  {author} {\bibinfo {author} {\bibfnamefont {K.~T.}\ \bibnamefont
  {Geier}}, \bibinfo {author} {\bibfnamefont {G.~I.}\ \bibnamefont {Martone}},
  \bibinfo {author} {\bibfnamefont {P.}~\bibnamefont {Hauke}}, \bibinfo
  {author} {\bibfnamefont {W.}~\bibnamefont {Ketterle}},\ and\ \bibinfo
  {author} {\bibfnamefont {S.}~\bibnamefont {Stringari}},\ }\href@noop {}
  \Eprint
  {https://arxiv.org/abs/2210.10064} {arXiv:2210.10064} \BibitemShut {NoStop}%
\bibitem [{\citenamefont {Lin}\ \emph {et~al.}(2011)\citenamefont {Lin},
  \citenamefont {{Jim{\'e}nez-Garc{\'i}a}},\ and\ \citenamefont
  {Spielman}}]{lin_Spinorbitcoupled_2011}%
  \BibitemOpen
  \bibfield  {author} {\bibinfo {author} {\bibfnamefont {Y.-J.}\ \bibnamefont
  {Lin}}, \bibinfo {author} {\bibfnamefont {K.}~\bibnamefont
  {{Jim{\'e}nez-Garc{\'i}a}}},\ and\ \bibinfo {author} {\bibfnamefont {I.~B.}\
  \bibnamefont {Spielman}},\ }\href@noop {} {\bibfield  {journal} {\bibinfo
  {journal} {Nature}\ }\textbf {\bibinfo {volume} {471}},\ \bibinfo {pages}
  {83} (\bibinfo {year} {2011})}\BibitemShut {NoStop}%
\bibitem [{\citenamefont {Li}\ \emph {et~al.}(2012)\citenamefont {Li},
  \citenamefont {Pitaevskii},\ and\ \citenamefont
  {Stringari}}]{li_Quantum_2012}%
  \BibitemOpen
  \bibfield  {author} {\bibinfo {author} {\bibfnamefont {Y.}~\bibnamefont
  {Li}}, \bibinfo {author} {\bibfnamefont {L.~P.}\ \bibnamefont {Pitaevskii}},\
  and\ \bibinfo {author} {\bibfnamefont {S.}~\bibnamefont {Stringari}},\
  }\href@noop {} {\bibfield  {journal} {\bibinfo  {journal} {Phys. Rev. Lett.}\
  }\textbf {\bibinfo {volume} {108}},\ \bibinfo {pages} {225301} (\bibinfo
  {year} {2012})}\BibitemShut {NoStop}%
\bibitem [{\citenamefont {Martone}\ \emph {et~al.}(2014)\citenamefont
  {Martone}, \citenamefont {Li},\ and\ \citenamefont
  {Stringari}}]{martone_Approach_2014}%
  \BibitemOpen
  \bibfield  {author} {\bibinfo {author} {\bibfnamefont {G.~I.}\ \bibnamefont
  {Martone}}, \bibinfo {author} {\bibfnamefont {Y.}~\bibnamefont {Li}},\ and\
  \bibinfo {author} {\bibfnamefont {S.}~\bibnamefont {Stringari}},\ }\href@noop
  {} {\bibfield  {journal} {\bibinfo  {journal} {Phys. Rev. A}\ }\textbf
  {\bibinfo {volume} {90}},\ \bibinfo {pages} {041604(R)} (\bibinfo {year}
  {2014})}\BibitemShut {NoStop}%
\bibitem [{\citenamefont {Zhang}\ \emph {et~al.}(2019)\citenamefont {Zhang},
  \citenamefont {Gui},\ and\ \citenamefont {Chen}}]{zhang_Nonlinear_2019}%
  \BibitemOpen
  \bibfield  {author} {\bibinfo {author} {\bibfnamefont {Y.}~\bibnamefont
  {Zhang}}, \bibinfo {author} {\bibfnamefont {Z.}~\bibnamefont {Gui}},\ and\
  \bibinfo {author} {\bibfnamefont {Y.}~\bibnamefont {Chen}},\ }\href@noop {}
  {\bibfield  {journal} {\bibinfo  {journal} {Phys. Rev. A}\ }\textbf {\bibinfo
  {volume} {99}},\ \bibinfo {pages} {023616} (\bibinfo {year}
  {2019})}\BibitemShut {NoStop}%
\bibitem [{\citenamefont {Wu}\ and\ \citenamefont
  {Niu}(2003)}]{wu_Superfluidity_2003}%
  \BibitemOpen
  \bibfield  {author} {\bibinfo {author} {\bibfnamefont {B.}~\bibnamefont
  {Wu}}\ and\ \bibinfo {author} {\bibfnamefont {Q.}~\bibnamefont {Niu}},\
  }\href@noop {} {\bibfield  {journal} {\bibinfo  {journal} {New J. Phys.}\
  }\textbf {\bibinfo {volume} {5}},\ \bibinfo {pages} {104} (\bibinfo {year}
  {2003})}\BibitemShut {NoStop}%
\bibitem [{\citenamefont {Li}\ \emph {et~al.}(2019)\citenamefont {Li},
  \citenamefont {Qu}, \citenamefont {Niffenegger}, \citenamefont {Wang},
  \citenamefont {He}, \citenamefont {Blasing}, \citenamefont {Olson},
  \citenamefont {Greene}, \citenamefont {{Lyanda-Geller}}, \citenamefont
  {Zhou}, \citenamefont {Zhang},\ and\ \citenamefont {Chen}}]{li_Spin_2019}%
  \BibitemOpen
  \bibfield  {author} {\bibinfo {author} {\bibfnamefont {C.-H.}\ \bibnamefont
  {Li}}, \bibinfo {author} {\bibfnamefont {C.}~\bibnamefont {Qu}}, \bibinfo
  {author} {\bibfnamefont {R.~J.}\ \bibnamefont {Niffenegger}}, \bibinfo
  {author} {\bibfnamefont {S.-J.}\ \bibnamefont {Wang}}, \bibinfo {author}
  {\bibfnamefont {M.}~\bibnamefont {He}}, \bibinfo {author} {\bibfnamefont
  {D.~B.}\ \bibnamefont {Blasing}}, \bibinfo {author} {\bibfnamefont {A.~J.}\
  \bibnamefont {Olson}}, \bibinfo {author} {\bibfnamefont {C.~H.}\ \bibnamefont
  {Greene}}, \bibinfo {author} {\bibfnamefont {Y.}~\bibnamefont
  {{Lyanda-Geller}}}, \bibinfo {author} {\bibfnamefont {Q.}~\bibnamefont
  {Zhou}}, \bibinfo {author} {\bibfnamefont {C.}~\bibnamefont {Zhang}},\ and\
  \bibinfo {author} {\bibfnamefont {Y.~P.}\ \bibnamefont {Chen}},\ }\href@noop
  {} {\bibfield  {journal} {\bibinfo  {journal} {Nat. Commun.}\ }\textbf
  {\bibinfo {volume} {10}},\ \bibinfo {pages} {375} (\bibinfo {year}
  {2019})}\BibitemShut {NoStop}%
\bibitem [{\citenamefont {Sun}\ and\ \citenamefont
  {Xie}(2005)}]{sun_Definition_2005}%
  \BibitemOpen
  \bibfield  {author} {\bibinfo {author} {\bibfnamefont {Q.-f.}\ \bibnamefont
  {Sun}}\ and\ \bibinfo {author} {\bibfnamefont {X.~C.}\ \bibnamefont {Xie}},\
  }\href@noop {} {\bibfield  {journal} {\bibinfo  {journal} {Phys. Rev. B}\
  }\textbf {\bibinfo {volume} {72}},\ \bibinfo {pages} {245305} (\bibinfo
  {year} {2005})}\BibitemShut {NoStop}%
\bibitem [{\citenamefont {Zhu}\ \emph {et~al.}(2015)\citenamefont {Zhu},
  \citenamefont {Sun},\ and\ \citenamefont {Wu}}]{zhu_Superfluidity_2015}%
  \BibitemOpen
  \bibfield  {author} {\bibinfo {author} {\bibfnamefont {Q.}~\bibnamefont
  {Zhu}}, \bibinfo {author} {\bibfnamefont {Q.-f.}\ \bibnamefont {Sun}},\ and\
  \bibinfo {author} {\bibfnamefont {B.}~\bibnamefont {Wu}},\ }\href@noop {}
  {\bibfield  {journal} {\bibinfo  {journal} {Phys. Rev. A}\ }\textbf {\bibinfo
  {volume} {91}},\ \bibinfo {pages} {023633} (\bibinfo {year}
  {2015})}\BibitemShut {NoStop}%
\bibitem [{\citenamefont {Zhang}\ \emph {et~al.}(2012)\citenamefont {Zhang},
  \citenamefont {Ji}, \citenamefont {Chen}, \citenamefont {Zhang},
  \citenamefont {Du}, \citenamefont {Yan}, \citenamefont {Pan}, \citenamefont
  {Zhao}, \citenamefont {Deng}, \citenamefont {Zhai}, \citenamefont {Chen},\
  and\ \citenamefont {Pan}}]{zhang_Collective_2012}%
  \BibitemOpen
  \bibfield  {author} {\bibinfo {author} {\bibfnamefont {J.-Y.}\ \bibnamefont
  {Zhang}}, \bibinfo {author} {\bibfnamefont {S.-C.}\ \bibnamefont {Ji}},
  \bibinfo {author} {\bibfnamefont {Z.}~\bibnamefont {Chen}}, \bibinfo {author}
  {\bibfnamefont {L.}~\bibnamefont {Zhang}}, \bibinfo {author} {\bibfnamefont
  {Z.-D.}\ \bibnamefont {Du}}, \bibinfo {author} {\bibfnamefont
  {B.}~\bibnamefont {Yan}}, \bibinfo {author} {\bibfnamefont {G.-S.}\
  \bibnamefont {Pan}}, \bibinfo {author} {\bibfnamefont {B.}~\bibnamefont
  {Zhao}}, \bibinfo {author} {\bibfnamefont {Y.-J.}\ \bibnamefont {Deng}},
  \bibinfo {author} {\bibfnamefont {H.}~\bibnamefont {Zhai}}, \bibinfo {author}
  {\bibfnamefont {S.}~\bibnamefont {Chen}},\ and\ \bibinfo {author}
  {\bibfnamefont {J.-W.}\ \bibnamefont {Pan}},\ }\href@noop {} {\bibfield
  {journal} {\bibinfo  {journal} {Phys. Rev. Lett.}\ }\textbf {\bibinfo
  {volume} {109}},\ \bibinfo {pages} {115301} (\bibinfo {year}
  {2012})}\BibitemShut {NoStop}%
\end{thebibliography}
\end{document}